\documentclass[journal]{IEEEtran}

\ifCLASSINFOpdf
\else
   \usepackage[dvips]{graphicx}
\fi
\usepackage{url}

\usepackage[switch]{lineno}

\usepackage{booktabs} 
\usepackage{float}
\usepackage{graphicx}
\usepackage{amsmath}
\usepackage{amsfonts}
\usepackage{caption}
\usepackage{subcaption}
\usepackage{multirow}
\usepackage{titlesec}
\usepackage{geometry}
\usepackage{fancyhdr}
\fancypagestyle{plain}{%
	\fancyhf{} 

	\fancyfoot[C]{\footnotesize\bfseries This is a preprint version for personal use only. The official version is available at: \url{https://ieeexplore.ieee.org/document/10816305}}
}

\titlespacing{\section}{0pt}{*0.9}{*0.9}
\titlespacing{\subsection}{0pt}{*0.5}{*0.39}
\begin{document}

\title{GALD-SE: Guided Anisotropic Lightweight Diffusion for Efficient Speech Enhancement}

\author{Chengzhong Wang, Jianjun Gu, Dingding Yao,
 Junfeng Li and Yonghong Yan, \IEEEmembership{Member, IEEE}
\thanks{
The authors are with the Key Laboratory of Speech Acoustics and
Content Understanding, Institute of Acoustics, Chinese Academy of
Sciences, University of Chinese Academy of Sciences, Beijing 100190,
China.}}

\maketitle

\begin{abstract}
Speech enhancement is designed to enhance the intelligibility and quality of speech across diverse noise conditions. Recently, diffusion model has gained lots of attention in speech enhancement area, achieving competitive results. Current diffusion-based methods blur the signal with isotropic Gaussian noise and recover clean speech from the prior. However, these methods often suffer from a substantial computational burden. We argue that the computational inefficiency partially stems from the oversight that speech enhancement is not purely a generative task; it primarily involves noise reduction and completion of missing information, while the clean clues in the original mixture do not need to be regenerated. In this paper, we propose a method that introduces noise with anisotropic guidance during the diffusion process, allowing the neural network to preserve clean clues within noisy recordings. This approach substantially reduces computational complexity while exhibiting robustness against various forms of noise and speech distortion. Experiments demonstrate that the proposed method achieves state-of-the-art results with only approximately 4.5 million parameters, a number significantly lower than that required by other diffusion methods. This effectively narrows the model size disparity between diffusion-based and predictive speech enhancement approaches. Additionally, the proposed method performs well in very noisy scenarios, demonstrating its potential for applications in highly challenging environments. The code is available at \url{https://github.com/wangchengzhong/GALDSE}.
\end{abstract}

\IEEEpeerreviewmaketitle

\section{Introduction}

\IEEEPARstart{S}{peech} enhancement aims at recovering clean speech signal from a noisy mixture that is corrupted by various types of noise. It has a wide range of applications, including hearing aids, speech recognition systems, and human-computer interaction. Traditional methods usually involve noise power estimation and extracting the clean signal in a statistical signal processing way \cite{tradiationalse1, tradiationalse2, tradiationalse3}. Not long ago, numerous machine learning-based speech enhancement techniques were proposed \cite{rnnoise, fullsubnet, dccrn, conv-tasnet}. These methods involve training a neural network to predict the clean signal from its noisy counterpart, significantly surpassing traditional methods in terms of speech quality, especially in non-stationary environments \cite{se_overview}. Recently, diffusion-based generative models have gained increasing attention \cite{sgmse+}--\cite{sbse}, achieving state-of-the-art results and surpassing predictive approaches when the tested audio mismatches with training datasets \cite{sgmse+}, \cite{bbed}.

Diffusion-based methods operate through a two-phase process: gradually adding noise to an initial state in the forward process, and then iteratively sampling along the reverse process to generate targeted samples. In the field of speech enhancement, the spectrogram of noisy speech frequently serves as a component of the prior for the diffusion process and as an input to the neural network for reverse sampling \cite{sgmse+}, \cite{sgmse}, \cite{cdiffuse}, \cite{vpidm}, or exclusively the latter \cite{universal, aaai_se}. The corresponding diffusion process can be interpreted in a discrete \cite{ddpm} or continuous \cite{sgmse+} way, while the latter permits the use of numerical stochastic differential equation (SDE) solvers to generate target speech \cite{fast-ode}. Richter et al. \cite{sgmse+} proposed an SDE-based method that introduces a novel drift term. In this way, the mean state of the process is transformed from clean speech into its noisy counterpart during the forward process and reversed during sampling. Two main approaches have been explored to improve upon this scheme: (\romannumeral 1) Carefully designing new interpolation techniques, such as the Brownian bridge \cite{se-bridge,bbed}, variance-preserving interpolation methods \cite{vpidm,vpidm-interspeech} and Schrödinger bridge \cite{sbse}, which have demonstrated significant performance improvements; and (\romannumeral 2) Integrating diffusion models with predictive methods \cite{storm, predictive decoders, thunder}. These methods can sometimes reduce the computational complexity \cite{thunder, sbse}, however, they're still highly computational-intensive.
We argue that this burden stems from the inefficiency of the diffusion process when applied to speech enhancement tasks. The uniform corruption fails to preserve the clean speech features in the mixture, resulting in a large distributional gap between diffusion states and clean speech, which may increase computational costs during inference.

Based on these insights, we propose GALD-SE, a Guided Anisotropic Lightweight Diffusion-based Speech Enhancement method, which incorporates anisotropic noise guidance in the diffusion process. The guidance effectively preserves the integrity of clean clues within noisy mixtures, enabling the process to directly access these clues at any timestep. The formulation of anisotropic diffusion is built upon the foundational work presented in \cite{resshift}. Experimental results demonstrate that our method achieves performance comparable to state-of-the-art while utilizing only 4.5 million parameters. Ablation analysis and comparison confirm the efficiency and superior quality of generated speech across various performance metrics.

\section{Proposed Method}
\label{jesus}
\begin{figure*}[h]
	\centering
	\begin{minipage}[b]{0.60\textwidth}
		\centering
		\includegraphics[width=\textwidth]{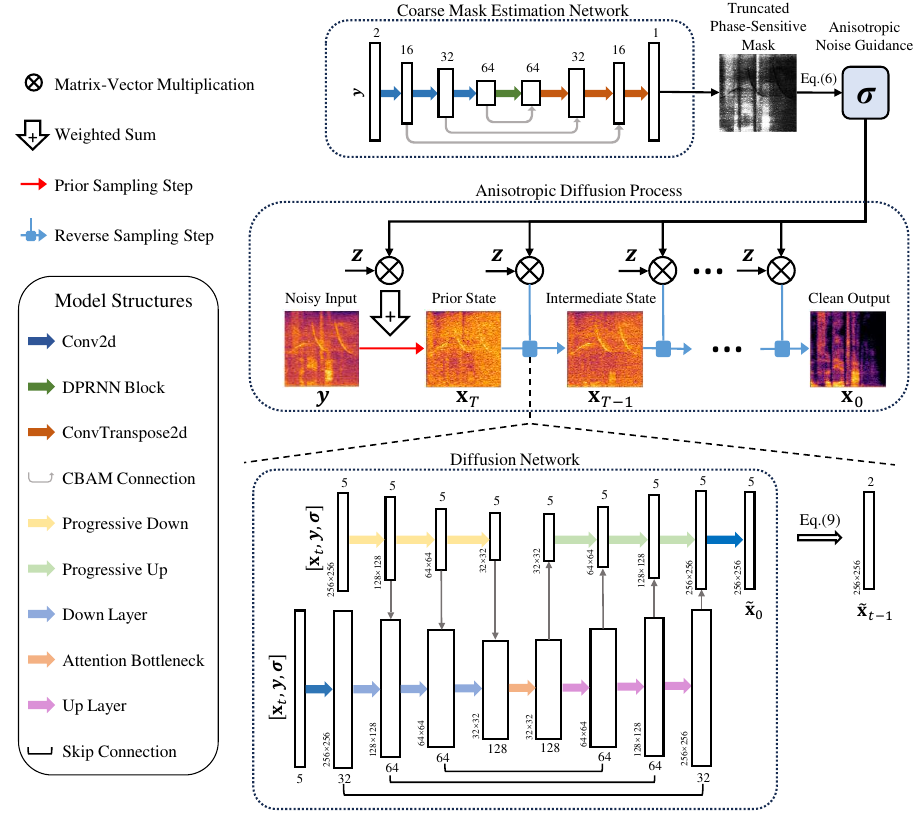}
		\captionsetup{font=small} 
\caption{System Overview. An anisotropic guidance is derived from a coarse mask estimation network and subsequently applied to the diffusion process. The prior state is sampled using noisy speech and guidance, after which the clean output is obtained by iteratively executing the reverse process, facilitated by the diffusion network. For detailed model structures, refer to \cite{sgmse+, nhs_sm_se}.}
		\label{fig:sys_overview}
	\end{minipage}
	\hfill
	\begin{minipage}[b]{0.36\textwidth}
		\vspace{0pt} 
		\begin{subfigure}[t]{0.99\textwidth}
			\centering
			\includegraphics[width=0.69\textwidth]{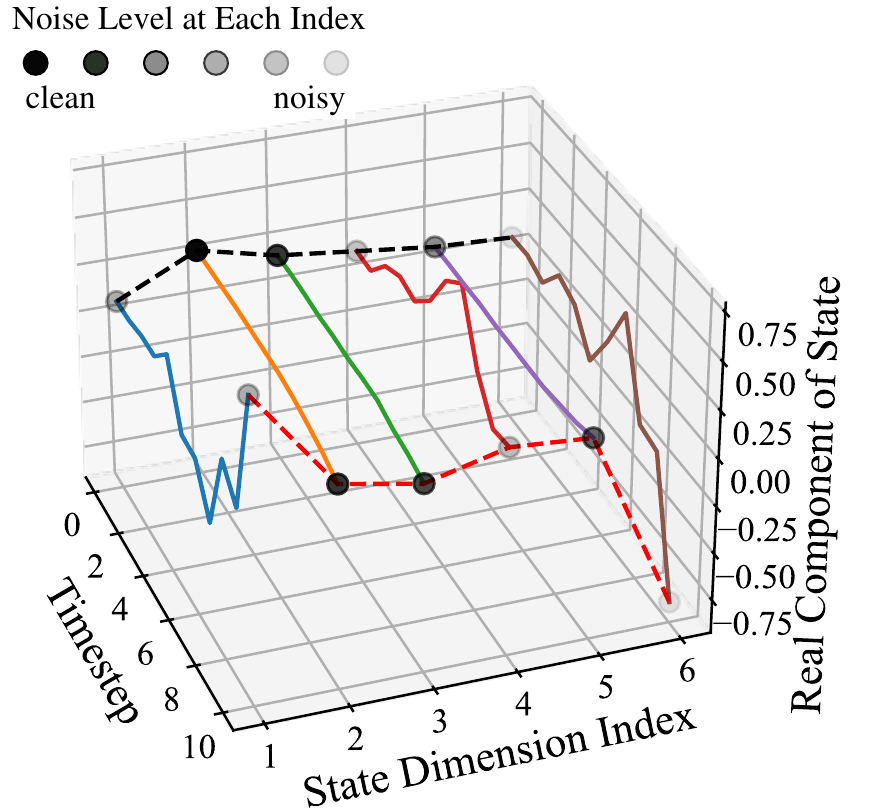}
			\captionsetup[subfigure]{font={footnotesize,rm}, labelfont={footnotesize,rm}}
			\caption{}
			\label{fig:toy_example}
		\end{subfigure}
		\\ 
		\begin{subfigure}[b]{0.99\textwidth}
			\centering
			\includegraphics[width=0.9\textwidth]{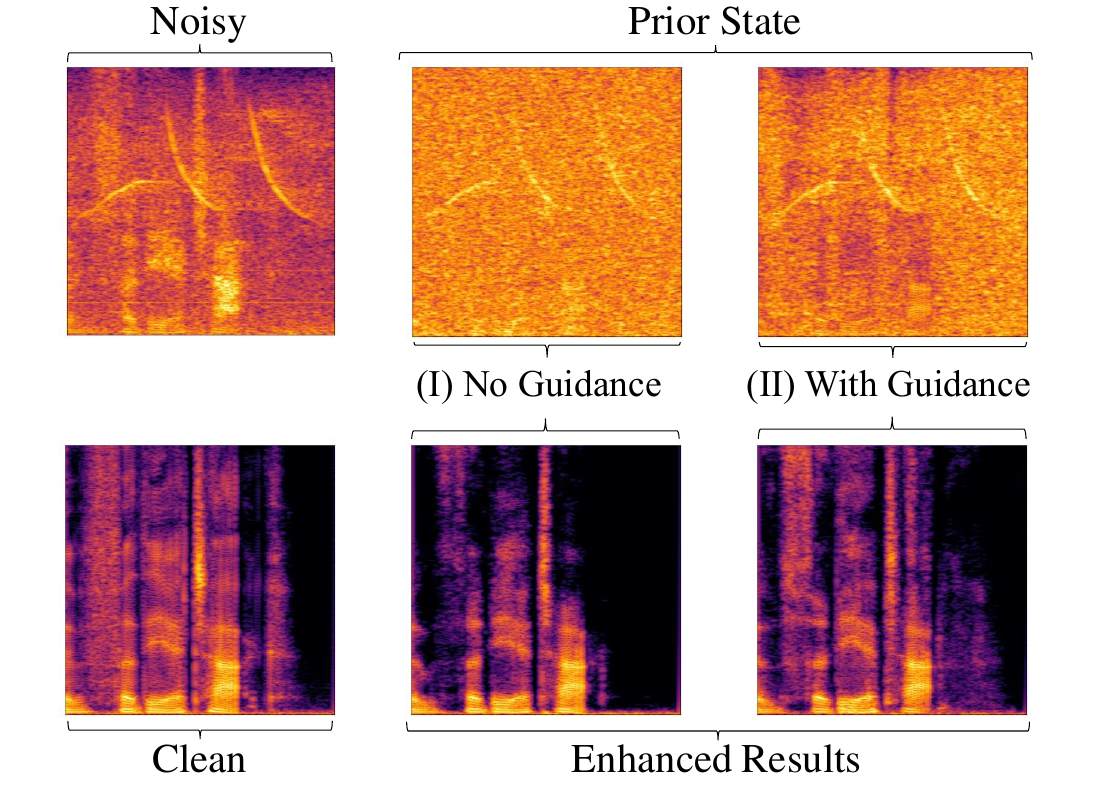}
			\captionsetup[subfigure]{font={footnotesize,rm}, labelfont={footnotesize,rm}}
			\caption{}
			\label{fig:spectrograms}
		\end{subfigure}
		\captionsetup{ font=small} 
		\caption{Visualization of anisotropic diffusion process and its impact on speech enhancement.
			(a) Illustrative example of the anisotropic diffusion process.
			 (b) Impact of anisotropic guidance on the prior state and final enhanced output.}
		\label{fig2}
	\end{minipage}
\end{figure*}

We begin with an overview of our proposed system, as illustrated in Fig. \ref{fig:sys_overview}. Our system enhances speech by leveraging the anisotropic diffusion process, with the anisotropic guidance derived from a coarse mask estimation. The system comprises three main components: (\romannumeral 1) a coarse mask estimation network for deriving guidance, (\romannumeral 2) a guided anisotropic diffusion process for selectively blurring noisy components, and (\romannumeral 3) a diffusion network for iterative refinement. The enhancement procedure begins by estimating the anisotropic guidance and sampling a prior state. It then proceeds through our guided reverse diffusion process facilitated by the diffusion network, ultimately yielding the enhanced clean speech.
While the core network architectures are adapted from existing models \cite{sgmse+, nhs_sm_se}, our key innovation lies in the anisotropic design of the diffusion process, which significantly reduces the computational load while achieving superior performance. The following subsections detail our main contributions: modifications to the original diffusion process, computation of anisotropic guidance, and specialized training and sampling strategies.
\subsection{Stochastic Process}
We denote the complex STFT feature vectors of clean and noisy speech as $\mathbf{x}_0$ and $\mathbf{y}$ respectively, both residing in $\mathbb{C}^d$ where $d=KF$, representing the coefficients of a flattened complex spectrogram. In the typical ``ResShift" formulation \cite{resshift}, during the forward process, the mean of the state gradually transitions from $\mathbf{x}_0$ to $\mathbf{y}$ by adding isotropic noise at each step. The conditional distribution of the state at time $t$, given the state at time $t-1$, the initial state $\mathbf{x}_0$, and the noisy feature $\mathbf{y}$, is given by:
\begin{equation}
	\begin{aligned}
		q(\mathbf{x}_t|\mathbf{x}_{t-1},\mathbf{x}_0, \mathbf{y})= \mathcal{N}_\mathbb{C}\left(\mathbf{x}_t;\mathbf{x}_{t-1} + \alpha_t (\mathbf{y} - \mathbf{x}_0), \kappa^2\alpha_t \mathbf{I}\right), \\ t= 1,...,T
	\end{aligned}
	\label{forward_orig}
\end{equation}
where $\mathcal{N}_\mathbb{C}$ denotes the circularly-symmetric complex normal distribution, $\mathbf{x}_t$ represents the state at timestep $t$, $\kappa$ is a hyperparameter controlling the noise variance, $\mathbf{I}\in \mathbb{R}^{KF\times KF}$ is the identity matrix, and $\alpha_t$ reflects the noise schedule.
We contend that the isotropic Gaussian noise added at each step can gradually impair the clean clues in the signal. Therefore, we propose to update each step of the forward process as follows:
\begin{equation}
	\begin{aligned}
	q(\mathbf{x}_t|\mathbf{x}_{t-1},\mathbf{x}_0, \mathbf{y})= \mathcal{N}_\mathbb{C}\left(\mathbf{x}_t,\mathbf{x}_{t-1} + \alpha_t (\mathbf{y} - \mathbf{x}_0), \kappa^2\alpha_t \boldsymbol{\sigma}^2\right) \\
	 t= 1,...,T
	\end{aligned}
	\label{forward}
\end{equation}
 where $\boldsymbol{\sigma}\in \mathbb{R}^{KF\times KF}$ is the anisotropic noise guidance matrix, designed to preserve the clean clues within noisy mixture. This matrix is diagonal, resulting in Gaussian noise that is independent across different T-F bins. During the forward process, the real and imaginary components of the complex features within each T-F bin undergo independent diffusion with the same variance, which is decided by the estimated proportion of noise in that bin. Fig. 2(a) illustrates the anisotropic process. For clarity, we focus on the real component of the state; the imaginary component follows a similar pattern. Suppose that in the noisy feature vector, indices 2, 3, and 5 contain a smaller proportion of noise and should therefore be blurred less, while indices 1, 4, and 6 contain a larger proportion of noise and should undergo more blurring. Consequently, for indices 2, 3, and 5, the Gaussian noise added at each step is relatively small, keeping the values at these positions nearly unchanged. In contrast, for the other indices, the Gaussian noise added at each timestep is much larger. In this way, the diffusion process can selectively blur T-F bins with varying intensities during the forward process, thereby preventing the introduction of excessive noise to bins that predominantly contain clean speech components.

Although the anisotropic guidance introduces varying intensities of noise to different bins, the overall mean of the process remains unaffected due to the zero-mean nature of the Gaussian noise. Consequently, following \cite{resshift}, we can directly sample $\mathbf{x}_t$ from the initial state $\mathbf{x}_0$ as follows:
\begin{equation}
	\begin{aligned}
	q\left(\mathbf{x}_t|\mathbf{x}_0, \mathbf{y}\right)=\mathcal{N}_\mathbb{C}\left(\mathbf{x}_t ; (1-\overline{\alpha}_t)\mathbf{x}_0+\overline{\alpha}_t \mathbf{y},\kappa^2\overline{\alpha}_t \boldsymbol{\sigma}^2 \right),\\ t = 1, ..., T
	\end{aligned}
	\label{sampling from initial}
\end{equation}
where $\overline{\alpha}_t =\Sigma_{k=1}^t\alpha_k$.
The corresponding reverse process can be depicted as:
\begin{equation}
	q\left(\mathbf{x}_{t-1} |  \mathbf{x}_t, \mathbf{x}_0, \mathbf{y}\right)=\mathcal{N}_\mathbb{C}\left(\mathbf{x}_{t-1} | (1-\beta_t) \mathbf{x}_t+ \beta_t \mathbf{x}_0, \overline{\boldsymbol{\sigma}}_t^2\right)
	\label{reverse step}
\end{equation}
where $\beta_t = \frac{\alpha_t}{\overline{\alpha}_t}$ and $\overline{\boldsymbol{\sigma}}_t^2 = \kappa^2 \beta_t (1-\beta_t){\boldsymbol{\sigma}}^2$. The calculation of the anisotropic guidance $\boldsymbol{\sigma}$ will be detailed in the following subsection.

\subsection{Anisotropic Guidance}\label{Anisotropic Guidance}

	We developed anisotropic guidance specifically to blur the noise structure while minimally altering the clean components. CMEN, denoted as $\mathbf{g}_\delta(\mathbf{y})$, is applied to derive this guidance. It is a lightweight UNet model\label{coarse mask network}  adapted from \cite{nhs_sm_se}, which employs a representative structure commonly used in speech enhancement. The network takes $\mathbf{y}$ as input to estimate the truncated phase-sensitive mask ${\mathbf{M}} \in \mathbb{R}^{K\times F}$ \cite{psm}, which is bounded within $[0,1]$:
\begin{equation}
	M_{k,f} = \text{clip}_{[0,1]}\left(\cos\theta_{k,f} \cdot \frac{|x_0|_{k,f}}{|y|_{k,f}}\right)
\end{equation}
where $\theta$ denotes the phase difference between $x_0$ and $y$. The phase-sensitive design aligns with our complex-domain diffusion process, providing a coarse estimation of speech proportion in each T-F bin. 
We use the estimated mask $\tilde{\mathbf{M}}:=\mathbf{g}_\delta(\mathbf{y})$ to define the guidance matrix $\boldsymbol{\sigma}$, which is diagonal, with its elements obtained as follows:
\begin{equation}
	\boldsymbol{\sigma} = \operatorname{diag}(\operatorname{vec}(\mathbf{1}-\tilde{\mathbf{M}}))
	\label{noise_mask_guidance}
\end{equation}
where $\mathbf{1}$ is a matrix in $\mathbb{R}^{K\times F}$ with all elements set to 1. The higher the noise ratio in a bin, the greater the corresponding value of $\sigma$ is in that bin. In this way, we encourage that the noise in the origin mixture loses its original structure during the forward process, while the clean structure remains nearly intact.

We further give an visualization for the impact of anisotropic noise guidance as in Fig. 2(b). The top row displays the noisy spectrogram, followed by the prior sampling states without and with guidance, respectively. The second row indicates the corresponding clean speech and the enhanced results. We observe that with anisotropic guidance, the T-F structure of the noise is significantly blurred, while the clean speech clues are preserved more effectively compared to its counterpart without guidance. This preservation facilitates the model's estimation of the original clean spectrogram. Consequently, the enhancement result with anisotropic guidance retains more speech details than those without guidance.
\subsection{Training and Sampling}
	We follow the ResShift \cite{resshift} formulation to perform training and sampling, albeit with the incorporation of anisotropic guidance.  The diffusion network, denoted as $\mathbf{f}_\theta(\mathbf{x}_t, \mathbf{y}, \boldsymbol{\sigma}, t)$, is parameterized by $\theta$ and processes inputs including the current state $\mathbf{x}_t$, noisy feature $\mathbf{y}$, anisotropic guidance $\boldsymbol{\sigma}$, and the current timestep $t$. The guidance $\boldsymbol{\sigma}$ is derived from the CMEN through (\ref{noise_mask_guidance}), allowing the diffusion network to adapt to potential estimation errors in CMEN.
At training stage, we design the diffusion model to estimate the initial state $\mathbf{x}_0$, as the presence of anisotropic guidance complicates the direct estimation of Gaussian noise. 	The CMEN is trained using the phase-sensitive spectrum approximation described in \cite{psm}. 
These two components are simultaneously trained using a mean square error criterion, leading to the following optimization objective:
\begin{equation}
	\min_{\theta,\delta} \sum_{t} \left[ \|\mathbf{f}_{\theta}(\mathbf{x}_t,\mathbf{y},\boldsymbol{\sigma},t) - \mathbf{x}_0\|_2^2 + \|\mathbf{g}_{\delta}(\mathbf{y})\odot \mathbf{y} - \mathbf{x}_0\|_2^2 \right]
\end{equation}
where $\odot$ denotes element-wise multiplication. It should be noticed that gradients from the diffusion network are not backpropagated to the CMEN, allowing each component to focus on its specific task.

At sampling stage, we first employ  (\ref{noise_mask_guidance}) to derive the anisotropic guidance $\boldsymbol{\sigma}$, and then get $\overline{\boldsymbol{\sigma}}_T$ based on (\ref{reverse step}). To initiate the reverse sampling procedure, we sample the prior state as follows:
\begin{equation}
	\mathbf{x}_T = \mathbf{y} +  \overline{\boldsymbol{\sigma}}_T\mathbf{z}, \quad \mathbf{z} \sim \mathcal{N}_{\mathbb{C}}(\mathbf{z} ; 0, \mathbf{I})
\end{equation}
At each reverse step $t$, we derive the previous state $\mathbf{x}_{t-1}$ using the estimated initial state $\tilde{\mathbf{x}}_0=\mathbf{f}_\theta(\mathbf{x}_t,\mathbf{y}, \boldsymbol{\sigma}, t)$:
\begin{equation}
	\mathbf{x}_{t-1} = (1-\beta_t) \mathbf{x}_t+ \beta_t \tilde{\mathbf{x}}_0 + \overline{\boldsymbol{\sigma}}_t \mathbf{z}, \quad \mathbf{z} \sim \mathcal{N}_{\mathbb{C}}(\mathbf{z} ; 0, \mathbf{I})
	\label{reverse_sampling}
\end{equation}
where $\beta_t$ and $\overline{\boldsymbol{\sigma}}_t$ retain their definitions as in (\ref{reverse step}). By applying this reverse sampling process, we iteratively derive the final estimation $\mathbf{\tilde{x}}_{0, \text{final}}$ from the prior state $\mathbf{x}_T$.
\section{Experiments}
	\begin{table*}[!htbp] 
	\centering
	\footnotesize
	\renewcommand{\arraystretch}{0.86}
	\captionsetup{font=footnotesize, skip=0pt}
	\caption{Performance comparison of different methods on DNS Simu Dataset}
	\scriptsize
	\setlength{\tabcolsep}{3pt}
	\begin{tabular}{l|ccc|cccc|cccc|cccc|cccc}
		\hline\hline
		\addlinespace[1.5pt]
		Metrics &\multirow{2}{*}{Type}& \multirow{2}{*}{Para.} & \multirow{2}{*}{Steps} & \multicolumn{4}{c|}{PESQ} & \multicolumn{4}{c|}{ESTOI} & \multicolumn{4}{c|}{SI-SNR (dB)} & \multicolumn{4}{c}{DNSMOS} \\
		\addlinespace[1pt]
		\cline{1-1}
		\cline{5-20}
		\addlinespace[1pt]
		SNR(dB)& & & &-15$\sim$-5 & -5$\sim$5 & 5$\sim$15 & Avg. & -15$\sim$-5 & -5$\sim$5 & 5$\sim$15 & Avg. & -15$\sim$-5 & -5$\sim$5 & 5$\sim$15 & Avg.& -15$\sim$-5 & -5$\sim$5 & 5$\sim$15 & Avg. \\
		\addlinespace[1.5pt]
		\hline
		\addlinespace[1pt]
		Noisy&-& - & - & 1.17 & 1.17 & 1.44 & 1.26 & 0.37 & 0.56 & 0.75 & 0.56 & -9.96 & -0.36 & 9.65 & -0.22 & 2.46 & 2.68 & 3.00 & 2.71\\
		\addlinespace[1pt]
		\hline
		\addlinespace[1pt]
		DEMUCS \cite{demucs} & D & 33.5M & - & 1.27 & 1.74 & 2.46 & 1.82 & 0.48 & 0.76 & 0.87 & 0.70 & 2.33 & 12.28 & 18.23 & 10.94 & 2.88 & 3.32 & 3.60 & 3.27\\
		CMGAN \cite{cmgan}& D&1.83M& - & 1.54 & 2.11 & 2.77 & 2.14 & 0.62 & 0.80 & 0.88 & 0.77 & 3.11 & 7.33 & 13.75 & 8.06 & 3.04 & 3.43 & 3.66 & 3.38\\
		DeepFilterNet \cite{dfnet2}&D&\textbf{1.80M}&-& 1.32 & 1.84&2.64& 1.93 & 0.50 & 0.74&0.85& 0.70 & 3.28 &10.50&16.43& 10.07& 2.79 & 3.22 & 3.61 & 3.21\\
		\addlinespace[1pt]  
		\hline
		\addlinespace[1pt]  
		SGMSE+ \cite{sgmse+}&G& 65M& 30*2& 1.38 & 1.99 & 2.78 & 2.05 & 0.53 & 0.80 & 0.88 & 0.74 & -1.43 & 11.10 & 18.12 & 9.27& 3.57 & 3.87 & 3.92 &3.79 \\
		StoRM \cite{storm}&D+G&55.6M & 15*2  & 1.43 & 1.96 & 2.73 & 2.04 & 0.59 & 0.80 & 0.88 & 0.76 & 3.97 & 12.23 & 18.63 & 11.61 & \textbf{3.80}& \textbf{3.92} & 3.94 & \textbf{3.89} \\
		VPIDM \cite{vpidm} & G & 65M & 25*2 & 1.55 & 2.13 & 2.88 & 2.19& 0.62 & \textbf{0.83} & 0.88& 0.78 & 6.17 & 13.18& 19.08 & 12.81 & 3.62 & 3.88 & \textbf{3.95} & 3.82\\
		BBED \cite{bbed}& G&65M & 30*2 & 1.58 & 2.14 & 2.91 & 2.21 & \textbf{0.65} & \textbf{0.83} & \textbf{0.89} & \textbf{0.79} & \textbf{7.34} & \textbf{14.13} & \textbf{19.94} & \textbf{13.81} & 3.58 & 3.84 & 3.92 & 3.78 \\
		GALD-SE (Ours)&D+G&4.5M&\textbf{6}& \textbf{1.61}& \textbf{2.23}& \textbf{3.02} & \textbf{2.29}& 0.63& \textbf{0.83} & \textbf{0.89}& 0.78& 6.83 & 13.42 & 19.32& 13.19 &3.69 &3.83& 3.92&3.81\\
		\addlinespace[1pt]  
		\hline\hline
	\end{tabular}
	\label{tab:comparison}
	\label{low_snr_table}
	\vspace*{-6pt}
\end{table*}
\subsection{Datasets and Evaluation Metrics}
We use DNS-Challenge \cite{dns2022} and VoiceBank+Demand (VBD) \cite{vbd} as the verified datasets. For the DNS corpus, we generated 1,000,000 pairs for training and 300 pairs for validation, each lasting 6 seconds, with signal-to-noise ratios (SNR) ranging from -15 to 25 dB. To construct the simulated test set, we randomly selected 1000 English read speech clips, distinct from those in the training set, and 1000 noise clips, each with a duration of 6 seconds. These clips were mixed at various SNR levels ranging from -15 dB to 15 dB, divided into three intervals\label{testdtb}.
The VBD corpus, widely used as a benchmark dataset in speech enhancement, consists of a training set with 28 speakers subjected to 8 noise types across 5 SNR levels, and a test set featuring 2 speakers with unseen noise types.
 
We choose the following widely used metrics to evaluate the performance: \textit{1) PESQ:} the Perceptional Evaluation of Speech Quality \cite{pesq}; \textit{2) ESTOI}: the Extended Short-Time Objective Intelligibility \cite{estoi}; \textit{3) SI-SNR}: Scale-Invariant Signal-to-Noise Ratio; \textit{4) CSIG, CBAK, COVL:} the speech quality (CSIG), background noise quality (CBAK) and the overall quality (COVL) of audio \cite{csig_cbak_covl}; \textit{5) DNSMOS:} A non-intrusive perceptual objective speech quality metric to evaluate noise suppressors \cite{dnsmos}.

\subsection{Hyperparameters and Training Settings}
	\label{hyperparameters}
For the stochastic process, we set the number of diffusion steps $T=6$ and the noise variance strength $\kappa=0.5$ in (\ref{forward}). The noise schedule $\overline{\alpha}_t$ is bounded between $\overline{\alpha}_1=0.001$ and $\overline{\alpha}_T=0.999$. A non-uniform geometric schedule is adopted for intermediate steps:
${\overline{\alpha}_t} = {\overline{\alpha}_1} \times 
\left({{\overline{\alpha}_T}/{\overline{\alpha}_1}}\right)^
{\left(\frac{t-1}{T-1}\right)^p}$, where $p=0.3$.
We adapted the diffusion network architecture from \cite{sgmse+}. As in Fig. 1, the first-step feature dimension is set to 32, resulting in 3.6 million parameters for the diffusion model. The CMEN, adapted from \cite{nhs_sm_se}, uses only a mask decoder, contributing 0.9 million parameters. Consequently, the entire system comprises 4.5 million parameters.

For the STFT configuration, we employ a 510-point FFT with a hop size of 128 samples. We compress the resulting spectrum using an exponential factor of 0.5 and then scale it by a factor of 0.5. For training, we set the batch size to 15 and use a learning rate of $10^{-4}$. The model is trained for 1000 epochs on the VBD corpus and 60 epochs on the DNS corpus.

\subsection{Results}

We initially compare our method with others using the simulated DNS corpus, as presented in Table \ref{low_snr_table}, where `D' and `G' denote discriminative and generative approaches, respectively. To ensure a comprehensive evaluation, we select representative methods from both categories, encompassing a range of established and state-of-the-art techniques. For the discriminative approach, we choose widely-adopted advanced methods, including DEMUCS \cite{demucs}, CMGAN \cite{cmgan}, and DeepFilterNet \cite{dfnet2}. Regarding the generative approach, we focus on recent diffusion-based methods, such as SGMSE+ \cite{sgmse+}, StoRM \cite{storm}, VPIDM \cite{vpidm} and BBED \cite{bbed}. Our results demonstrate that the proposed method achieves performance comparable to state-of-the-art techniques across multiple evaluation metrics. Although our method shows a slight performance gap in ESTOI and SI-SNR metrics compared to the best-performing BBED approach, it achieves this with significantly fewer parameters (4.5M vs. 65M) and sampling steps (6 vs. 60). We attribute this significant computational efficiency to our novel implementation of anisotropic noise guidance, which preserves clean speech clues and streamlines the diffusion process. It is worth noting that PESQ may lack reliability in low-SNR scenarios, where ESTOI emerges as a more suitable metric. Our results demonstrate that our method achieves an ESTOI rank of No. 2 among all methods, despite having significantly fewer parameters than the top-ranked BBED method. This underscores the efficacy of our approach in challenging acoustic environments. 
	\vspace{0pt}
\begin{table}[!htbp] 
	\centering
	\renewcommand{\arraystretch}{0.87}
	\captionsetup{font=footnotesize}
	\caption{Performance Comparison of Different Methods on VoiceBank/DEMAND Dataset}
	\scriptsize
	\begin{tabular}{@{}l@{\hspace{3pt}}|@{\hspace{3pt}}c@{\hspace{3pt}}c@{\hspace{3pt}}c@{\hspace{3pt}}|@{\hspace{3pt}}c@{\hspace{3pt}}c@{\hspace{3pt}}c@{\hspace{3pt}}c@{\hspace{3pt}}c@{}}
		\toprule
		Methods  & Type & Para. & Steps & PESQ & ESTOI  & CSIG & CBAK & COVL \\ \midrule
		Mixture  & -& - & - & 1.97 & 0.79 &  3.35 & 2.44 & 2.63 \\ \midrule
		NCSN++ \cite{score model}& D&  3.6M & -  & 2.73 & 0.86 & 3.45 & 3.39 & 3.17 \\
		UNet-Small \cite{nhs_sm_se}& D & \textbf{0.9M} & - & 2.85 & 0.86 & 3.97 & 3.43 & 3.52 \\
		DEMUCS \cite{demucs} &D&  33.5M & -  & 3.07 & - &  \textbf{4.31} & 3.40 & 3.63 \\
		\midrule
		CDiffuSE \cite{cdiffuse}&G& 10.1M  &\textbf{6}& 2.52 & - &  3.72 & 2.91 & 3.10 \\
		SGMSE+ \cite{sgmse+} &G& 65M & 30*2 & 2.93 & 0.87 & 4.12 & 3.37 & 3.51 \\
		StoRM \cite{storm}& D+G&55.6M  & 15*2 &2.93 & \textbf{0.88} &  - & - & -\\
		Thunder \cite{thunder} &D+G& 27.8M & 30*2& 2.97 & 0.87 & - & - & - \\
		VPIDM \cite{vpidm}  &G& 65M & 25*2  & 3.16 & 0.87 & 4.23 & 3.53 & 3.70 \\
		\midrule
		GALD-SE (Ours) &  D+G&4.5M & \textbf{6}  & \textbf{3.19} & \textbf{0.88} & 4.23 & \textbf{3.61} & \textbf{3.72}\\ 
		\quad --reduced CMEN  & D+G & 3.9M & \textbf{6} & 3.15 & \textbf{0.88} & 4.21 & 3.57 & \textbf{3.72} \\ 
		\quad --with isotropic noise &D+G& 4.5M&\textbf{6}&2.93 & 0.87& 4.10 & 3.26 & 3.50\\	
		\quad --noise-free&D+G&4.5M&\textbf{6} & 2.80 & 0.83 & 4.02 & 3.11 & 3.38\\
		\bottomrule
	\end{tabular}
	\vspace*{-15pt}
	\label{vbd}
\end{table}
\vspace{0pt}

Subsequently, we compare our method with others on the VBD dataset.  The results are presented in Table \ref{vbd}. We include the recently proposed Thunder \cite{thunder} in this evaluation, as it was originally assessed on the VBD dataset, enabling direct performance comparison. The NCSN++ approach adapts the diffusion model architecture to function as a discriminative backbone, whereas the UNet approach exclusively employs the first-stage coarse mask estimation, leaving the phase unaltered.
Our results demonstrate that the combination of diffusion with anisotropic guidance outperforms approaches using either the coarse mask frontend or diffusion as a discriminative backbone alone. Moreover, when compared to other state-of-the-art methods, our proposed approach achieves superior performance across most of the metrics.
We posit that this advantage arises from our novel strategy of deeply integrating clean speech clues into the diffusion process through guidance, which enables a more nuanced and effective speech enhancement process.

Finally, we conduct an ablation study to demonstrate the effectiveness of our anisotropic guidance method, as shown in the last four rows of Table \ref{vbd}. 
We evaluate three variants: the first adopting a reduced CMEN architecture with two encoder-decoder layers, the second incorporating isotropic noise during diffusion, and the third implementing a noise-free process. The last case corresponds to the cold diffusion formulation \cite{cold_diffusion}. For fair comparison, all variants incorporate the discriminative guidance $\boldsymbol{\sigma}$ as input to the diffusion network. Results from the first variant show only marginal performance degradation with reduced CMEN parameters, demonstrating our framework's robustness. The isotropic noise and noise-free variants, however, exhibit evident performance deterioration, validating the necessity of guided anisotropic noise in our method.
\section{Conclusion}
This letter introduces a novel diffusion-based speech enhancement method that leverages guided anisotropic noise. By focusing on the preservation of clean speech clues in noisy environments, our approach reduces the computational demands typically associated with diffusion-based methods, and achieves superior performance. Further reduction of computational load and exploration of the potential for real-time audio stream processing are currently under investigation.


\begin{thebibliography}{}
\bibitem{tradiationalse1}S. Boll, ``Suppression of acoustic noise in speech using spectral subtraction,'' {\em IEEE Transactions on acoustics, speech, and signal processing}, vol. 27, no. 2, pp. 113--120, 1979.
\bibitem{tradiationalse2}M. Dendrinos, S. Bakamidis, and G. Carayannis, ``Speech enhancement from noise: A regenerative approach,'' {\em Speech Communication}, vol. 10, no. 1, pp. 45--57, 1991.
\bibitem{tradiationalse3}Y. Ephraim and H. L. Van Trees, ``A signal subspace approach for speech enhancement,'' {\em IEEE Transactions on speech and audio processing}, vol. 3, no. 4, pp. 251--266, 1995.
\bibitem{rnnoise}J. M. Valin, ``A hybrid DSP/deep learning approach to real-time full-band speech enhancement,'' in {\em 2018 IEEE 20th international workshop on multimedia signal processing}, 2018, pp. 1--5.
\bibitem{fullsubnet}X. Hao, X. Su, R. Horaud, and X. Li, ``Fullsubnet: A full-band and sub-band fusion model for real-time single-channel speech enhancement,'' in {\em ICASSP}, 2021, pp. 6633--6637.
\bibitem{dccrn}Y. Hu, Y. Liu, S. Lv, M. Xing, S. Zhang, Y. Fu, J. Wu, B. Zhang, and L. Xie, ``DCCRN: Deep complex convolution recurrent network for phase-aware speech enhancement,'' in {\em Proc. Interspeech}, 2020, pp. 2472–2476.
\bibitem{conv-tasnet}Y. Luo and N. Mesgarani, ``Conv-tasnet: Surpassing ideal time–frequency magnitude masking for speech separation,'' {\em IEEE/ACM transactions on audio, speech, and language processing,} vol. 27, no. 8, pp. 1256--1266, 2019.
\bibitem{se_overview}C. Zheng, H. Zhang, W. Liu, X. Luo, A. Li, X. Li, and B. C. Moore, ``Sixty years of frequency-domain monaural speech enhancement: From traditional to deep learning methods,'' {\em Trends in Hearing}, vol. 27, Art. no. 23312165231209913, 2023.
\bibitem{sgmse+} J. Richter, S. Welker, J.-M. Lemercier, B. Lay, and T. Gerkmann, ``Speech enhancement and dereverberation with diffusion-based generative models,'' {\em IEEE/ACM Transactions on Audio, Speech, and Language Processing }, vol. 31, pp. 2351--2364, 2023. 

\bibitem{sgmse}S. Welker, J. Richter, and T. Gerkmann, ``Speech enhancement with score-based generative models in the complex STFT domain,'' in {\em Proc. Interspeech}, 2022, pp. 2928–2932.

\bibitem{cdiffuse}Y.-J. Lu, Z.-Q. Wang, S. Watanabe, A. Richard, C. Yu, and Y. Tsao, ``Conditional diffusion probabilistic model for speech enhancement,'' in {\em ICASSP}, 2022, pp. 7402--7406.
\bibitem{storm}J.-M. Lemercier, J. Richter, S. Welker, and T. Gerkmann, “StoRM: A diffusion-based stochastic regeneration model for speech enhancement and dereverberation,'' {\em IEEE/ACM Transactions on Audio, Speech, and Language Processing}, pp. 2724--2737, 2023.
\bibitem{vpidm}  Z. Guo, Q. Wang, J. Du, J. Pan, Q. -F. Liu and C. -H. Lee, ``A variance-preserving interpolation approach for diffusion models with applications to single channel speech enhancement and recognition,'' {\em IEEE/ACM Transactions on Audio, Speech, and Language Processing}, vol. 32, pp. 3025--3038, 2024.
\bibitem{vpidm-interspeech} Z. Guo, J. Du, C. -H. Lee,  Y. Gao,  W. Zhang, ``Variance-preserving-based interpolation diffusion models for speech enhancement,'' in {\em Proc. Interspeech}, 2023, pp. 1065--1069.
\bibitem{se-bridge}Z. Qiu, M. Fu, F. Sun, G. Altenbek, and H. Huang, ``Se-bridge: Speech enhancement with consistent brownian bridge,''  2023, {\em  arXiv:2305.13796}.
\bibitem{bbed}B. Lay, S. Welker, J. Richter, and T. Gerkmann, ``Reducing the prior mismatch of stochastic differential equations for diffusion-based speech enhancement,'' in {\em Proc. Interspeech,} 2023, pp. 3809--3813.
\bibitem{dose}W. Tai, Y. Lei, F. Zhou, G. Trajcevski, and T. Zhong, ``Dose: Diffusion dropout with adaptive prior for speech enhancement,'' {\em Advances in Neural Information Processing Systems}, vol. 36, pp. 40272--40293, 2024.
\bibitem{universal} J. Serrà, S. Pascual, J. Pons, R. O. Araz, and D. Scaini,  ``Universal speech enhancement with score-based diffusion,'' 2022, {\em arXiv:2206.03065}.
\bibitem{casual_diffusion_se}J. Richter, S. Welker, J. -M. Lemercier, B. Lay, T. Peer, and T. Gerkmann, ``Causal Diffusion Models for Generalized Speech Enhancement,'' {\em IEEE Open Journal of Signal Processing}, vol. 5, pp. 780--789, 2024.
\bibitem{predictive decoders}H. Shi, K. Shimada, M. Hirano, T. Shibuya, Y. Koyama, Z. Zhong, S. Takahashi, T. Kawahara, and Y. Mitsufuji, ``Diffusion-based speech enhancement with joint generative and predictive decoders,'' in {\em ICASSP}, 2024, pp. 12951--12955.
\bibitem{pretrain_guided_diffusion} Y. Yang, N. Trigoni, and A. Markham, ``Pre-training feature guided diffusion model for speech enhancement,'' 2024, {\em arXiv:2406.07646}.
\bibitem{aaai_se}W. Tai, F. Zhou, G. Trajcevski, and T. Zhong, ``Revisiting denoising diffusion probabilistic models for speech enhancement: condition collapse, efficiency and refinement,'' in {\em Proceedings of the AAAI Conference on Artificial Intelligence}, 2023, pp. 13627--13635.
\bibitem{thunder}T. Trachu, C. Piansaddhayanon, and E. Chuangsuwanich, ``Thunder: Unified regression-diffusion speech enhancement with a single reverse step using brownian bridge,'' 2024, {\em  arXiv:2406.06139}.
\bibitem{sbse}A. Jukić, R. Korostik, J. Balam, and B. Ginsburg, ``Schrödinger Bridge for Generative Speech Enhancement,'' in {\em Proc. Interspeech}, 2024, pp. 1175--1179.
\bibitem{score model}Y. Song, J. Sohl-Dickstein, D.P. Kingma, A. Kumar, S. Ermon, and B. Poole, ``Score-based generative modeling through stochastic differential equations,'' in {\em Proc. ICLR}, 2021, pp. 1--36.
\bibitem{fast-ode}Z. Zhou, D. Chen, C. Wang, and C. Chen, ``Fast ode-based sampling for diffusion models in around 5 steps,'' in {\em Proceedings of the IEEE/CVF Conference on Computer Vision and Pattern Recognition}, 2024, pp. 7777--7786.
\bibitem{cold_diffusion}H. Yen, F. G. Germain, G. Wichern and J. L. Roux, ``Cold Diffusion for Speech Enhancement,'' in {\em ICASSP}, 2023, pp. 1--5.
\bibitem{resshift}Z. Yue, J. Wang, and C. C. Loy, ``Resshift: Efficient diffusion model for image super-resolution by residual shifting,''  in {\em Advances in Neural Information Processing Systems}, 2023, pp. 13294--13307.

\bibitem{ddpm}J. Ho, A. Jain, and P. Abbeel, ``Denoising diffusion probabilistic models,'' in {\em Advances in Neural Information Processing Systems}, 2020, pp. 6840--6851.

\bibitem{nhs_sm_se}W. Jiang and K. Yu, ``Speech enhancement with integration of neural homomorphic synthesis and spectral masking,'' {\em IEEE/ACM Transactions on Audio, Speech, and Language Processing}, vol. 31, pp. 1758--1770, 2023.
\bibitem{psm}H. Erdogan, J. R. Hershey, S. Watanabe and J. Le Roux, ``Phase-sensitive and recognition-boosted speech separation using deep recurrent neural networks," in {\em ICASSP}, 2015, pp. 708--712.
\bibitem{dns2022} H. Dubey, V. Gopal, R. Cutler, S. Matusevych, S. Braun, E. S. Eskimez, M. Thakker, T. Yoshioka, H. Gamper, and R. Aichner, ``ICASSP 2022 deep noise suppression challenge,'' in {\em ICASSP}, 2022, pp. 9271--9275.
\bibitem{vbd}C. Valentini-Botinhao, X. Wang, S. Takaki, and J. Yamagishi, ``Investigating RNN-based speech enhancement methods for noise-robust Text-to-Speech,'' in {\em Speech Synthesis Workshop}, 2016, pp. 146--152.
\bibitem{pesq}A. W. Rix, J. G. Beerends, M. P. Hollier and A. P. Hekstra, ``Perceptual evaluation of speech quality (PESQ)--a new method for speech quality assessment of telephone networks and codecs,'' in {\em ICASSP}, 2001, pp. 749--752. 
\bibitem{estoi}J. Jensen and C. H. Taal, ``An algorithm for predicting the intelligibility of speech masked by modulated noise maskers,'' {\em IEEE/ACM Transactions on Audio, Speech, and Language Processing,} vol. 24, no. 11, pp. 2009--2022, 2016.
\bibitem{csig_cbak_covl}Y. Hu and P. C. Loizou, ``Evaluation of Objective Quality Measures for Speech Enhancement,'' {\em IEEE Transactions on Audio, Speech, and Language Processing}, vol. 16, no. 1, pp. 229--238, 2008.
\bibitem{dnsmos}C. K. Reddy, V. Gopal, and R. Cutler, ``DNSMOS: A non-intrusive perceptual objective speech quality metric to evaluate noise suppressors,'' in {\em ICASSP}, 2021, pp. 6493--6497.
\bibitem{demucs} A. Defossez, G. Synnaeve, and Y. Adi, ``Real time speech enhancement in the waveform domain,'' in {\em Proc. Interspeech}, 2020, pp. 3291--3295.
\bibitem{cmgan} R. Cao, S. Abdulatif, and B. Yang, ``CMGAN: Conformer-based metric GAN for speech enhancement,'' in {\em Proc. Interspeech}, 2022, pp. 936--940.
\bibitem{dfnet2} H. Schröter, A. N. Escalante-B., T. Rosenkranz, and A. Maier, ``DeepFilterNet2: Towards Real-Time Speech Enhancement on Embedded Devices for Full-Band Audio,'' in {\em 17th International Workshop on Acoustic Signal Enhancement}, 2022, pp. 1--5.









\end{thebibliography}
\end{document}